\documentclass[
showpacs,preprintnumbers,amsmath,amssymb]{revtex4}
\usepackage{amsmath}
\usepackage{graphicx}
\begin{document}

\tolerance=5000

\def\be{\begin{equation}}
\def\ee{\end{equation}}
\def\bea{\begin{eqnarray}}
\def\eea{\end{eqnarray}}
\def\tr{{\rm tr}\, }
\def\nn{\nonumber \\}
\def\e{{\rm e}}

\title{$F(R)$ cosmology in presence of a phantom fluid and its scalar-tensor counterpart: towards a unified precision model of the universe evolution}

\author{Emilio Elizalde\footnote{E-mail: elizalde@ieec.uab.es, elizalde@math.mit.edu} and
Diego S\'aez-G\'omez\footnote{E-mail: saez@ieec.uab.es} }
\medskip

\affiliation{Consejo Superior de Investigaciones Cient\'{\i}ficas ICE/CSIC and IEEC \\
Campus  Universitat Aut\`onoma de Barcelona, Facultat de Ci\`encies \\
 Torre C5-Parell-2a planta, 08193 Bellaterra (Barcelona) Spain}

\begin{abstract}
The behavior of cosmological evolution is studied in the case when a phantom fluid, that contributes to the universe accelerated expansion, is introduced in an $F(R)$ model. At the early stages of the universe history, the dark fluid is seen to give rise to a deceleration of its expansion. For $t$ close to present time it works as an additional contribution to the effective cosmological constant and, later, it produces the transition to a phantom era, which could actually be taking place right now in some regions of the cosmos, and might have observable consequences. For $t$ close to the Rip time, the universe  becomes completely dominated by the dark fluid, whose equation of state is phantom-like at that time. Our model, which is able to reproduce the dark energy period quite precisely, may still be modified in such a way that the epoch dominated by an effective cosmological constant---produced by the $F(R)$ term and by the dark fluid contribution---becomes significantly shorter, what happens when a matter term is included. The dark fluid with phantom behavior gives rise to a super-accelerated phase, as compared with the case where just the viable $F(R)$ term contributes. It is also seen explicitly that an $F(R)$ theory can be constructed from a phantom model in a scalar-tensor theory, in which the scalar field does not behave as phantom (in the latter case the action for $F(R)$ would be complex). Promising $F(R)$ models which are able to cross the phantom divide in a convenient way are constructed explicitly.
\end{abstract}


\maketitle

\section{Introduction}


Our first aim in this paper will be to build a reliable cosmological model by using, as starting point, modified gravity theories of the family of the so-called $F(R)$ theories which comprise the class of viable models, e.g., those having an $F(R)$ function such that the theory can pass all known local gravity tests (see \cite{PowerR} -\cite{f(R)GRtests}).
In the next section, we will investigate cosmological evolution as coming from $F(R)$ gravity. We will consider, for $F(R)$, some candidates to produce inflation and cosmic acceleration in a unified fashion. In particular, we investigate in detail the behavior of $F(R)$ gravity as the contribution of a perfect fluid. As a crucial novelty, an additional matter fluid will be included which may play, as we shall see, quite an important cosmological role. In fact, it may decisively contribute to the two accelerated epochs of the Universe, what is to say that we will study a model where dark energy consists of two separate contributions. The possibilities to obtain precision cosmology are enhanced in this way.

Two main cases will be discussed: $F(R)$ cosmology with a constant equation of state (EoS) fluid, and
$F(R)$ cosmology in the presence of a phantom fluid. In the last case, a couple of specific
example will be worked through in detail, namely one with a phantom fluid with constant EoS and, as a second example, a fluid with dynamical EoS of the type proposed in Refs.~\cite{Darkfluids1} and \cite{Darkfluids2}. In this case a dark fluid is present which has an inhomogeneous EoS that can depend on the proper evolution of the Universe, what opens a number of very interesting possibilities.

As is well known, $F(R)$ gravity can be written in terms of a scalar field---quintessence or phantom like---by redefining the function $F(R)$ with the use of a scalar field, and then performing a conformal transformation.  It has been shown that, in general, for any given $F(R)$ the corresponding scalar-tensor theory can, in principle, be obtained, although the solution is going to be very different from one case to another. Also attention will be paid to the reconstruction of $F(R)$ gravity from a given scalar-tensor theory. It is known \cite{F(R)vsScalar2} that the phantom case in scalar-tensor theory does not exist, in general, when starting from $F(R)$ gravity. In fact, the conformal transformation becomes complex when the phantom barrier is crossed, and therefore the resulting $F(R)$ function becomes complex. We will see that, to avoid this hindrance, a dark fluid can be used in order to produce the phantom behavior in such a way that the $F(R)$ function reconstructed from the scalar-tensor theory continues to be real.
We will prove, in an explicit manner, that an $F(R)$ theory can indeed be constructed from a phantom model in a scalar-tensor theory, but where the scalar field does not behave as a phantom field (in which case the action for $F(R)$ would be complex). Moreover, we will explicitly show that very interesting and quite simple $F(R)$ models crossing the phantom divide can be constructed.

\section{Framework}
Our first aim in this paper will be to construct reliable cosmological models by using, as starting point, modified gravity theories of the family of the so-called $F(R)$ theories which comprise the class of  viable models, e.g., those having an $F(R)$ function such that the theory can pass all known local gravity tests. We now consider the action corresponding to one of these theories which, aside from the gravity part, also contains a matter contribution, namely
\be
S=\int d^4x \sqrt{-g}\left[R+F(R)+L_m \right]\ .
\label{1.1}
\ee
Here $\kappa^2=8\pi G$, and $L_m$ stands for the Lagrangian corresponding to matter of some kind. Note that the first term in (\ref{1.1}) is just the usual Hilbert-Einstein action, and the $F(R)$ term can be considered, as will be shown below, as the dynamical part of some kind of perfect fluid (see \cite{MyF(R)1} and \cite{F(R)vsScalar1AndEoS}), which may constitute an equivalent to the so-called dark fluids. The field equations corresponding to action (\ref{1.1}) are obtained by variation of this action with respect to the metric tensor $g_{\mu\nu}$, what yields
\be
R_{\mu\nu} -\frac{1}{2}g_{\mu\nu}R +  R_{\mu\nu} F(R)- \frac{1}{2} g_{\mu\nu} F(R) + g_{\mu\nu}  \Box F'(R) -  \nabla_{\mu} \nabla_{\nu}F'(R)=\kappa^2T_{\mu\nu}\ .
\label{1.2}
\ee
Here the primes denote derivatives with respect to $R$. We  assume a flat FRW metric, then the Friedmann equations are obtained as the 00 and the ij components. They take the  form
\bea
3H^2=\kappa^2 \rho_m - \frac{1}{2}F(R)+3(H^2+\dot{H})F'(R)-18F''(R)(H^2\dot{H}+H\ddot{H})\ , \nn
-3H^2-2\dot{H}=\kappa^2p_m+\frac{1}{2}F(R)-(3H^2+\dot{H})F'(R)- \Box F'(R)-HF''(R)\dot{R}\ ,
\label{1.3}
\eea
where $H(t)=\dot{a}/a$. Note that both equations are written in such a way that the $F(R)$-terms are put on the matter side; thus we may define an energy density and a pressure density for these $F(R)$-terms, as follows
\bea
\rho_{F(R)}=-\frac{1}{2}F(R)+ 3(H^2+\dot{H})F'(R)-18F''(R)(H^2\dot{H}+H\ddot{H})\ , \nn
p_{F(R)}=\frac{1}{2}F(R)-(3H^2+\dot{H})F'(R)- \Box F'(R)-HF''(R)\dot{R}\ .
\label{1.4}
\eea
Then, the Friedmann equations (\ref{1.3}) take a simple form, with two fluids contributing to the scale factor dynamics. From the energy and pressure densities defined in (\ref{1.4}), one can obtain the EoS for the dark fluid, defined in terms of the $F(R)$ components. This is written as
\bea
w_{F(R)}= \frac{\frac{1}{2}F(R)-(3H^2+\dot{H})F'(R)- \Box F'(R)-HF''(R)\dot{R}}{-\frac{1}{2}F(R)+ 3(H^2+\dot{H})F'(R)+\Box F'(R)-\nabla_0\nabla^0F'(R)} \nn
\longrightarrow \qquad p_{F(R)}= - \rho_{F(R)}+2\dot{H}F'(R)-\nabla_0\nabla^0F'(R)-HF''(R)\dot{R}
\ .
\label{1.5}
\eea
The EoS, Eq.~(\ref{1.5}), defines a fluid that depends on the Hubble parameters and its derivatives, so it may be considered as a fluid with inhomogeneous EoS (see \cite{Darkfluids1} or \cite{Darkfluids2}). In absence of any kind of matter, the dynamics of the Universe are carried out by the $F(R)$-component, which may be chosen so that it reproduces (or at least contributes) to the early inflation and  late-time acceleration epochs. In order to avoid serious problems with known physics, one has to choose the $F(R)$ function in order that the theory contains flat solutions and passes also the local gravity tests (see \cite{F(R)UnfInfCosAcce2andsolartest}-\cite{f(R)viable10}). To reproduce the whole history of the Universe, the following conditions on the $F(R)$ function have been proposed (see \cite{F(R)UnfInfCosAcce2andsolartest}-\cite{F(R)UnfInfCosAcce3}):
\begin{enumerate}
\renewcommand{\labelenumi}{\roman{enumi}}
\item) Inflation occurs  under one of the following conditions:
\be
\lim_{R\rightarrow\infty} F(R)=-\Lambda_i
\label{1.6}
\ee
or
\be
\lim_{R\rightarrow\infty} F(R)=  \alpha R^n\ .
\label{1.7}
\ee
In the first situation (\ref{1.6}), the $F(R)$ function behaves as an effective cosmological constant at early times, while the second condition yields accelerated expansion where the scale factor behaves as $a(t)\sim t^{2n/3}$.
\item) In order to reproduce late-time acceleration, we can impose on function $F(R)$ a condition similar to the one above. In this case the Ricci scalar has a finite value, which is assumed to be the current one, so that the condition is expressed as
\be
F(R_0)=-2R_0 \quad F'(R_0)\sim 0\ .
\label{1.8}
\ee
\end{enumerate}
Hence, under these circumstances, the $F(R)$ term is able to reproduce the two different accelerated epochs of the universe history. An interesting example that satisfies these conditions has been proposed in Ref. \cite{f(R)viable1}:
\be
F(R)=\frac{\mu^2}{2\kappa^2}\frac{c_1(\frac{R}{c_1})^k+c_3}{c_2(\frac{R}{\mu^2})^k+1}\ .
\label{1.8a}
\ee
This model is studied in detail in Ref.~\cite{f(R)deSitter}, where it is proven that the corresponding  Universe solution goes, in its evolution, through two different De Sitter points, being one of them stable and the other unstable and which can be identified as corresponding to the current accelerated era and to the inflationary epoch, respectively. Thus, the model (\ref{1.8a}) is able to reproduce both accelerated epochs when the free parameters are conveniently chosen---as was done in Ref.~\cite{f(R)deSitter}---in such way that the the F(R) theory in (\ref{1.8a}) produces two de Sitter epochs and grateful exit from the inflationary stage is achieved. In the same way, from a similar example given in Ref.~\cite{F(R)UnfInfCosAcce3}, we can consider the function
\be
F(R)=\frac{R^n(\alpha R^n-\beta)}{1+\gamma R^n}.
\label{1.9}
\ee
This function, represented in Fig.~1, gives rise to inflation at the early stages of the Universe, assuming the condition (\ref{1.7}) holds, while for the current epoch it behaves as an effective cosmological constant. This is explicitly seen in Fig.~1, where function (\ref{1.9}) is represented for the specific power $n=2$.

\begin{figure}
 \centering
 \includegraphics[width=3in,height=2in]{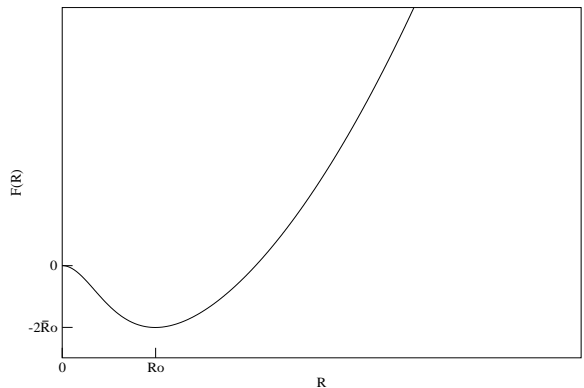}
 \caption{The function $F(R)$ as given by (\ref{1.9}) for $n=2$. We see that at the current epoch ($R\sim R_0$), $F(R)$ behaves as a cosmological constant while for $R\rightarrow\infty$ (inflation) its values grow as a power law.}
 \label{fig1}
\end{figure}

Function (\ref{1.9}) leads, at the current epoch,  to a perfect-fluid behavior with an EoS given by $p_{F(R)}\sim -\rho_{F(R)}$. In the next section such kind of $F(R)$ functions will be considered, with the inclusion of a matter Lagrangian in the action (\ref{1.1}), and the corresponding cosmological evolution will be studied. We will see that both inflation and the current acceleration can indeed be produced by the $F(R)$ fluid, provided other components are allowed to contribute too.

\section{Cosmological evolution from viable $F(R)$ gravity with a fluid}

In this section we will discuss cosmological evolution as coming from $F(R)$-gravity. We consider the function $F(R)$ as given by  (\ref{1.9}). Functions of this kind have been shown to yield viable models which comply with all known local gravity tests (see, e.g., \cite{f(R)viable10}), and they are good candidates to produce  inflation and cosmic acceleration in a unified fashion. We will study in detail the behavior of $F(R)$-gravity, by considering it as the contribution of a perfect fluid in the way already explained in the preceding section. As a crucial novelty, an additional matter fluid will be here incorporated, which may play, as we shall see, an important cosmological role. It may decisively contribute to the two accelerated epochs of the Universe, what is to say that we study a model where dark energy consists of two separate contributions. Some examples of phantom evolution will be then discussed, in which the $F(R)$ contribution acts as a cosmological constant and the additional fluid behaves as a phantom field, what opens again interesting new venues.

\subsection{$F(R)$ cosmology with a constant EoS fluid $p_m=w_m\rho_m $}

We consider a Universe governed by action (\ref{1.1}), where the $F(R)$ function is given by (\ref{1.9}). The matter term is represented by a perfect fluid, whose equation of state is $p_m=w_m\rho_m $ (where $w_m= $cons.) In this case,  by considering this $F(R)$ term as a perfect fluid---with energy and pressure densities given by (\ref{1.4})---the Friedmann equations reduce to Eqs.~(\ref{1.3}). For simplicity, we will study the case where $n=2$; then, the function $F(R)$ of our specific model reads (the study can be extended to other values of $n$ without much problem)
\be
F(R)= \frac{R^2(\alpha R^2-\beta)}{1+\gamma R^2}\ .
\label{2.1}
\ee
First of all, let us explain qualitatively what the aim of this model is. For simplicity, we neglect the contribution of matter, so the first Friedmann equation yields
\[3H^2=-\frac{R^2(\alpha R^2-\beta)}{2(1+\gamma R^2)}+3(H^2+\dot{H})\frac{2R(\alpha\gamma R^4-2\alpha R^2-\beta)}{(1+\gamma R^2)^2}-
 \]
\be
-18F''(R)(H^2\dot{H}+H\ddot{H})\frac{2\alpha\gamma^2 R^6+20\alpha\gamma R^4+6(\beta\gamma-\alpha)R^2-2\beta}{(1+\gamma R^2)^3}\ ,
\label{2.1a}
\ee
where $R=6(2H^2+\dot{H})$. It can be rewritten as a dynamical system (see Ref.~\cite{f(R)deSitter}):
\be
\dot{H}=C\ , \quad \dot{C}=F_1(H, C)\ ,
\label{2.1b}
\ee
and it can be shown that its critical points are those which give a constant Hubble rate ($\dot{H}=0$), i.e., the points that yield a de Sitter solution of the Friedmann equations. We can now investigate the existence of these points for the model (\ref{2.1}) by introducing the critical points $H=H_0$ in the Friedmann equation (\ref{2.1a})
\be
3H^2_0=-\frac{R^2(\alpha R^2-\beta)}{2(1+\gamma R^2)}+3H^2_0\frac{2R(\alpha\gamma R^4-2\alpha R^2-\beta)}{(1+\gamma R^2)^2}\ .
\label{2.1c}
\ee
To simplify, it can be rewritten in terms of the Ricci scalar $R_0=12H^2_0$, what yields
\be
\frac{\gamma}{4}R_0^5-\gamma\beta R_0^4+\frac{\gamma}{2}R_0^3+\frac{1}{4}R_0=0\ .
\label{2.1d}
\ee
This can be solved, so that the viable de Sitter points (positive roots) are found explicitly. A very simple study of Eq.~(\ref{2.1d}), by using Descartes' rule of signs, leads to the conclusion that Eq.~(\ref{2.1d}) can have either two or no positive roots. As shown below, one of these roots is identified as an effective cosmological constant that produces the current accelerated expansion of the Universe, while the second root can produce the inflationary epoch under some initial conditions. Then, the model described by the function (\ref{2.1}) is able to unify the expansion history of the Universe. In order to get a grateful exit from the inflationary epoch, stability in the vicinity of the critical points needs to be studied: the corresponding de Sitter point during inflation must be unstable. This can be achieved, as is already known for the function (\ref{1.8a}), by choosing specific values of the free parameters. Even in the case of stable dS inflation, exit from it can be achieved by coupling it with matter, by the effect of a small non-local term (or by some other mechanism).

Let us now study the details of this same model at the current epoch, when it is assumed that F(R) produces the cosmic acceleration and where the matter component is taken into account.
Function (\ref{2.1}) is depicted in Fig.~1; its minimum is attained at $R=R_0$, which is assumed to be the current value of the Ricci scalar. Further, $F(R)$ as given by (\ref{2.1}) behaves as a cosmological constant at  present time. Imposing the condition $\beta\gamma/\alpha>>1$  on the otherwise free parameters, the values of $R_0$ and $F(R_0)$ are then given by  \cite{F(R)UnfInfCosAcce3}
\be
R_0 \sim \left( \frac{\beta}{\alpha\gamma}\right)^{1/4}\ , \qquad F'(R_0)=0\ , \qquad  F(R_0)= -2\tilde{R_0}\sim -\frac{\beta}{\gamma}\ .
\label{2.2}
\ee
For simplicity, we shall study  the cosmological evolution around the present value of the Ricci scalar, $R=R_0$, where (\ref{2.1}) can be expressed as $F(R)=-2\tilde{R_0}+f_0(R-R_0)^2+O\left( (R-R_0)^3\right)$, the solution for the Friedmann equations (\ref{1.3}) can be written as $H(t)=H_0(t)+\delta H$, where at zero order the solution  is the same as in the case of a cosmological constant, namely
\be
H(t)= \sqrt{\frac{\tilde{R_0}}{3}} \coth\left( \frac{(1+w_m)\sqrt{3\tilde{R_0}}}{2}t\right)\ .
\label{2.3}
\ee
As pointed out in Ref.~\cite{F(R)UnfInfCosAcce3}, the perturbations $\delta H$ around the current point $R=R_0$ may be neglected. Therefore, we can study the evolution of the energy density (\ref{1.4}) for the $F(R)$ term as the EoS parameter defined by (\ref{1.5}) around the minimum of the $F(R)$ function, which is assumed to be the present value of the Ricci scalar. For such purposes, it is useful to rewrite the Hubble function (\ref{2.3}) as a function of the redshift $z$, instead of $t$. The relation between both variables can be expressed as
\be
\frac{1}{1+z}= \frac{a}{a_0}= \left[ A \sinh\left( \frac{(1+w_m)\sqrt{3\tilde{R_0}}}{2}\, t\right)\right]^{\frac{2}{3(1+w_m)}}\ ,
\label{2.4}
\ee
where $a_0$ is taken as the current value of the scale parameter, and $A^2=\rho_{0m}a^{-3(1+w_m)}_0$, being $\rho_{0m}$ the current value of the energy density of the matter contribution. Hence, the Hubble parameter (\ref{2.3}) is expressed  as a function of the redshift $z$, as
\be
H^2(z)=\frac{\tilde{R_0}}{3}\left[ A^2(1+z)^{3(w_m+1)}+1\right]\ .
\label{2.5}
\ee
Thus, the model characterized by the action (\ref{1.1}) with the function (\ref{2.1}) and a matter fluid with constant EoS, depends on the free parameters ($\alpha,\beta,\gamma$) contained in the expression of $F(R)$ and also on the value of the EoS parameter ($w_m$). When imposing the minimum value for the function $F(R)$ to take place at present time ($z=0$), the free parameters can be adjusted with the observable data. Then we study the behavior of our model close to $z=0$, where the contributions of non-linear terms produced by the function (\ref{2.1}) are assumed not to modify the solution (\ref{2.5}). In spite of the Hubble parameter being unmodified, for $z$ close to zero, the energy density $\rho_{F(R)}$ will in no way remain constant for small values of the redshift. To study the behavior of the $F(R)$ energy density given by (\ref{1.4}) it is convenient to express it as the cosmological parameter, $\Omega_{F(R)}=\frac{\rho_{F(R)}}{3H^2(z)}$, which can be written as
\be
\Omega_{F(R)}(z)=-\frac{F(R)}{6H^2(z)}+\left[ 1+\frac{\dot{H}(z)}{H^2(z)}\right] F'(R)-18F''(R)\left[ \dot{H}(z)+\frac{\ddot{H}(z)}{H(z)}\right]\ ,
\label{2.6}
\ee
where the Ricci scalar is given by $R=6[2H^2(z)+\dot{H}(z)]$. This expression (\ref{2.6}) can be studied as a function of the redshift. As we are considering the solution (\ref{2.5}), which has been calculated near $z=0$, where the $F(R)$ function has a minimum, the second term in the expression (\ref{2.6}) is negligible as compared with the other two terms ($F''(R_0)R_0/F'(R_0)\sim\sqrt{\beta\gamma/\alpha}>>1$). This aproximation, as the solution (\ref{2.5}), is valid for values of the Ricci scalar close to $R_0$, where the higher derivatives of $F(R)$ are small compared with the function. The aproximation is no longer valid when the F(R) derivatives are comparable with F(R). Then, as discussed above, the free parameters of the model can be fitted by the current observational values of the cosmological parameters, and by fixing the minimum of $F(R)$ to occur for $z\sim 0$.

\begin{figure}[h]
 \centering
 \includegraphics[width=4in,height=3in]{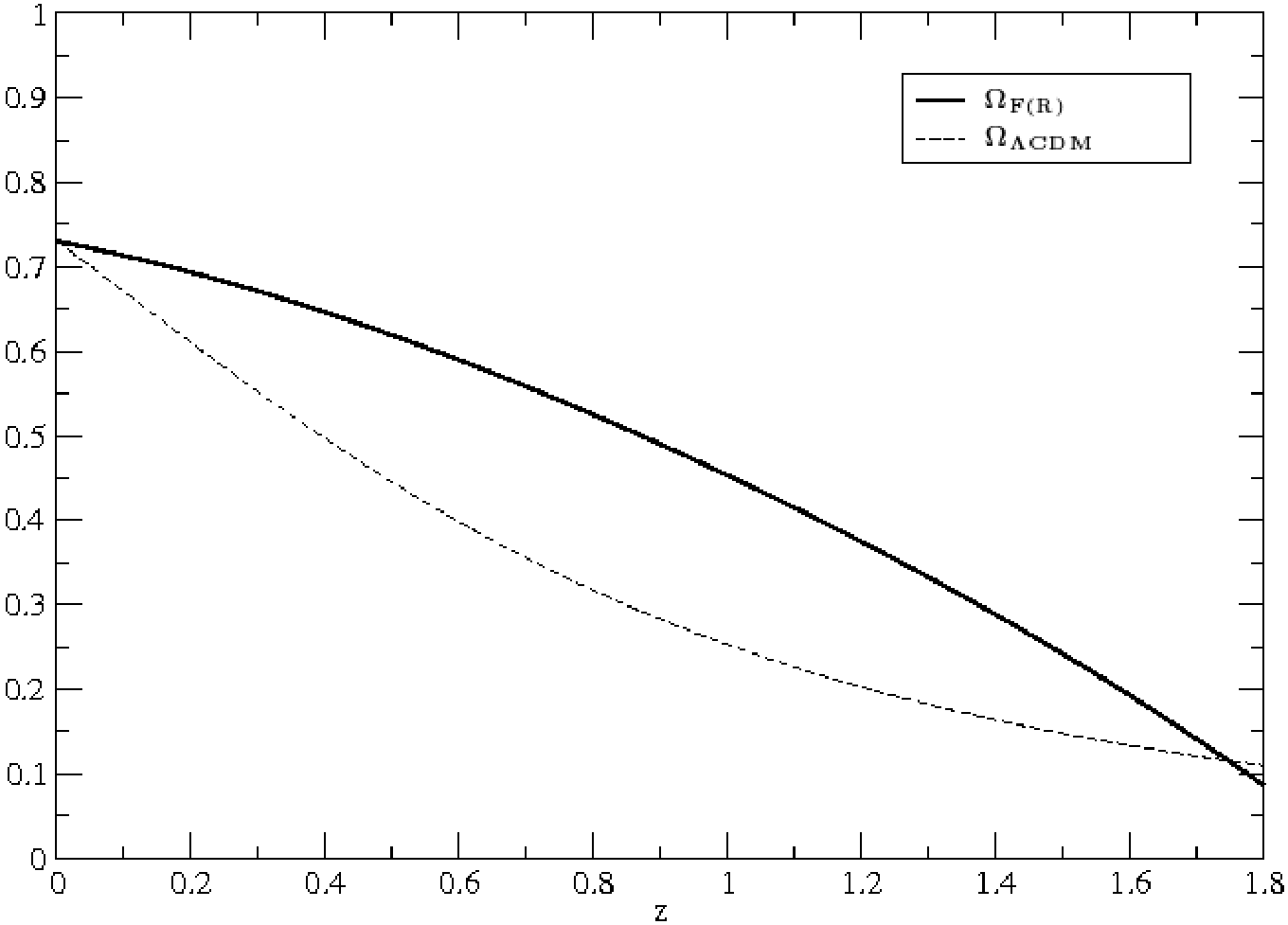}
 \caption{Evolution of the cosmological parameter from dark energy versus redshift, such for $F(R)$ theory as for $\Lambda CDM$ model.}
 \label{fig2}
\end{figure}

We shall use the value for the Hubble parameter $H_0=100\, h \ \text{km}\ \text{s}^{-1}\ \text{Mpc}^{-1}$ with $h=0.71\pm 0.03$ and the matter density $\Omega_m^0=0.27 \pm 0.04$ given in Ref.~\cite{WmapData}. In Fig.~\ref{fig2}, the evolution of the $F(R)$ energy density (\ref{2.6}) is represented for the model described by (\ref{2.1}), where the matter content is taken to be pressureless ($w_m=0$). The cosmological evolution shown corresponds to redshifts from  $z=1.8$ to $z=0$. For redshifts larger than $z=1.8$  perturbations around the solution (\ref{2.3}) are non-negligible, and the expression for the Hubble parameter (\ref{2.3}) is no more valid. However, in spite of the fact that the evolution shown in Fig.~\ref{fig2} is not a complete picture of the $F(R)$ model given by (\ref{2.1}), it still provides an illustrative example to compare it to the standard $\Lambda CDM$ model, which is also represented in Fig.~\ref{fig2}, around the present time. As shown in Fig.~2, both models have two common points for $z=0$ and  $z=1.74$, while the behavior of each model between such points is completely different from one another. This result shows the possible differences between $F(R)$ models of the type (\ref{1.9}) and the $\Lambda CDM$ model, although probably other viable models for $F(R)$ gravity may give a different adjustment to the $\Lambda CDM$ model (see \cite{f(R)viable1}-\cite{f(R)viable10}). Furthermore, the effective EoS parameter for the $F(R)$ fluid, $w_{F(R)}$, given by the expression (\ref{1.5}), is plotted in Fig.~3, again as a function of redshift. It is shown there that $w_{F(R)}$ is close to $-1$ for $z=0$, where the $F(R)$ fluid behaves like an effective cosmological constant, while it grows for redshifts up  to $z=1.5$, where it reaches a zero value. According to the analysis of observational dataset from Supernovae (see Ref.\cite{Supernova}), the results obtained for the evolution of the EoS parameter, represented in Fig.~3, are allowed by the observations.
\begin{figure}[h]
 \centering
 \includegraphics[width=4in,height=3in]{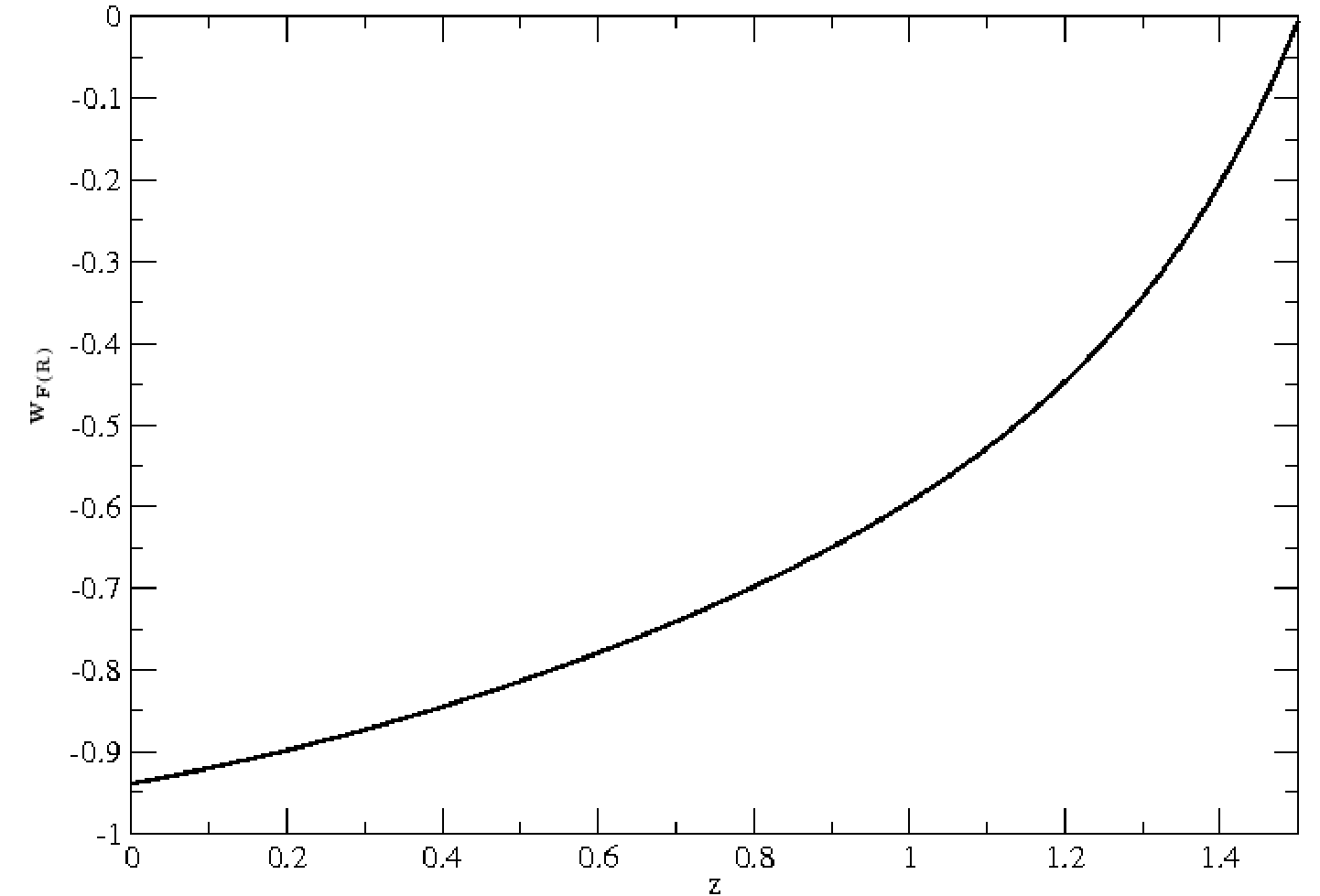}
 \caption{Effective EoS parameter $w_{F(R)}$ versus redshift. It takes values close to $-1$ for $z=0$, and it grows for higher redshifts.}
 \label{fig3}
\end{figure}

As a consequence, the $F(R)$ model given by (\ref{2.1}), and where  the $F(R)$ fluid behaves as an effective cosmological constant, is able indeed to reproduce the same behavior at present time as the $\Lambda CDM$ model. On the other hand, as was pointed out in the section above, the $F(R)$ model given by (\ref{2.1}) also reproduces the accelerated expansion of the inflation epoch, so that the next natural step to undertake with this kind of models should be to study their complete cosmological histories, and the explicit details allowing for a grateful exit from inflation, what should demonstrate their true potential. Also note that, in order to obtain a realistic well behaved model, further analysis should be carried out, as the comparison with the luminosity distance from Supernovae, or the data from CMB surveys, although this is a major task, that will be left for future work. Furthermore, $F(R)$ functions of this kind might even lead to a solution of the cosmological constant problem, by involving a relaxation mechanism of the cosmological constant, as was proposed in Ref.~\cite{CosmoProblem}. The effective cosmological constant obtained could eventually adjust to the observable value. This will also require a deeper investigation.

\subsection{$F(R)$ cosmology in presence of a phantom fluid}

According to several analysis of observational data (see \cite{obsData1} and \cite{obsData2}), the effective EoS parameter of the physical theory that governs our Universe should quite probably be less than $-1$, what means that we should be ready to cope with phantom behavior. This possibility has been explored in $F(R)$ theory \cite{F(R)vsScalar2}, where the possibility to construct an $F(R)$ function that reproduces this kind of behavior has been demonstrated, and where the possible future singularities envisaged, as the Big Rip, so common in phantom models, do take place. An example is given by (see \cite{MyF(R)1})
\be
f(R)\sim R^{1-\alpha/2}\ , \quad \text{where} \quad \alpha=\frac{H_0+1-\sqrt{H_0(H_0+10)+1}}{2}\ ,
\label{2.6a}
\ee
here $H_0$ is a positive constant, and $f(R)=R+F(R)$. In this case, the solution, neglecting other possible components, is the following
\be
H(t)\sim\frac{H_0}{t_s-t}\ ,
\label{2.6b}
\ee
where the Hubble parameter diverges at the Rip time $t_s$. Another kind of $F(R)$ gravity that reproduces a phantom behavior, but avoids the Big Rip singularity, is
\be
f(R)=\left[R-6H_0(H_0+1)+5/2) \right]\frac{R-2H_0\pm\sqrt{R(1-2H_0)}}{2}\ ,
\label{2.6c}
\ee
where $H_0$ is a positive constant (for other $F(R)$ models with transient phantom behavior, see \cite{F(R)vsScalar3}). The solution obtained is
\be
H(t)=H_0 t+ \frac{1}{2t}\ .
\label{2.6d}
\ee
In this last example, the universe has a phantom behavior at large time but there is no Big Rip singularity. However, the main problem of those models is that the corresponding $F(R)$ function is not well constrained at small scales.

We will  now study the behavior of the cosmological evolution when a phantom fluid is introduced that contributes to the accelerated expansion. A phantom fluid can be described, in an effective way, by an EoS $p_{ph}=w_{ph}\rho_{ph}$, where $w_{ph}<-1$. This EoS can be achieved, for example, with a negative kinetic term or in the context of scalar-tensor theory. The microphysical study of this kind of fluids is a problem of fundamental physics that has been studied in several works (see for example \cite{Phantom1} and \cite{Phantom2}), as it is also the study of  stability of the  solutions. But it is not the aim for this paper to discuss, in this depth, the features of phantom fields, and it will suffice for our purpose to consider the effective EoS that describes a phantom fluid.  The $F(R)$ function we will consider is the same as above, given  by (\ref{1.9}), which will contribute as an effective cosmological constant at present time. The motivation for this case study comes, as will be further explored, from the difficulty one encounters in constructing {\it viable} $F(R)$ functions which produce phantom epochs, and this problem has a translation in the scalar-tensor picture in the Einstein frame too. As recent observational data suggest (see \cite{obsData1} and \cite{obsData2}), the effective EoS parameter for dark energy is around $-1$, so that the phantom case is allowed (and even still favored) by recent, accurate, and extensive observations.\medskip

\noindent{\sf Example 1.} First, we consider a phantom fluid with constant EoS, e.g., $p_{ph}=w_{ph}\rho_{ph}$, where $w_{ph}<-1$ is a constant. As in the case above,  the $F(R)$ function will be given by (\ref{2.1}), and it will be assumed that its current value is attained at the minimum $R_0=\left(\beta/\alpha\gamma\right)^{1/4}$. We here neglect other contributions, as the dust term studied previously. Friedmann equations take the form
\be
H^2=\frac{\kappa^2}{2}\rho_{ph}+\frac{\tilde{R}_0}{3}\ , \quad \dot{H}=-\frac{\kappa^2}{3}\rho_{ph}(1+w_{ph})\ .
\label{2.7}
\ee
An expanding solution (note that a contracting solution can be obtained from the above equations too) for these equations is given by
\be
H(t)=\frac{3}{2|1+w_{ph}|(t_s-t)}+\sqrt{\frac{\tilde{R_0}}{3}}\ ,
\label{2.8}
\ee
where $t_s$ is called the Rip time, that means the instant where a future Big Rip singularity will take place. Although for $t$ much bigger than the present time, the solution (\ref{2.8}) is not valid anymore ---because perturbations, due to the derivatives of the function $F(R)$, become large--- and as the Ricci scalar $R=6(2H^2+\dot{H})$ grows with time, this model  will behave as $F(R)\sim R^2$ for large times, which is known to produce accelerated expansion, and whose behavior is  described by (see \cite{PowerR})
\be
H(t)\propto \frac{h_0}{t_s-t}\ ,
\label{2.8a}
\ee
where $h_0=\frac{4}{3|w_{ph}+1|}$. And the effective EoS parameter for large time is
\be
w_{eff}=-1-\frac{4}{3h_0}\ .
\label{2.8b}
\ee

 That is to say, the universe will go through a super-accelerated expansion stage due to the phantom fluid and to the contribution coming from $F(R)$, until it reaches the Big Rip singularity. In this case where no other contribution, as dust matter or radiation, is taken into account, late-time acceleration comes from the phantom behavior when the dark fluid component dominates, while the Universe behaves as dark energy when the $F(R)$ term is the dominant one (as was in absence of this phantom fluid). In other words, the universe would not accelerate as a phantom one alone, but it will as a dark energy fluid with $w_{eff}\geq -1$. This is a fundamental point: the additional dark fluid is essential for a universe described by the $F(R)$ function given in (\ref{2.1}) to display a phantom transition.

We can study the evolution of this phantom fluid by using the continuity equation:
\be
\dot{\rho}_{ph}+3H \rho_{ph}(1+w_{ph})=0\ ,
\label{2.9}
\ee
which can be solved, at the present time, when the Hubble parameter is expressed by Eq.~(\ref{2.8}). Then, the following solution is obtained
\be
\rho_{ph}=\rho_{0ph}\frac{\e^{|1+w_{ph}|\sqrt{3\tilde{R_0}}t}}{(t_s-t)^{9/2}}\ ,
\label{2.10}
\ee
where $\rho_{0ph}$ is an integration constant. As we can see, the energy density for the phantom fluid grows with time until the Rip value is reached where the energy density becomes infinite. On the other hand, the evolution of the $F(R)$ term may be studied qualitatively by observing the expression given for the energy density of this fluid in (\ref{1.4}), so that this evolution is similar to the one in the above case. That is, as the value of the Ricci scalar increases with time, it is natural to suppose that, in the past, the energy density belonging to $F(R)$  had smaller values than at present, so that the matter dominated epoch could occur when the $F(R)$ fluid and the phantom fluid were much less important than they are now. Again, for this kind of model the $F(R)$ contribution amounts currently to an effective cosmological constant which drives the universe's acceleration. \medskip

\noindent{\sf Example 2.} As a second example of a phantom fluid, we consider one with a dynamical EoS of the type proposed in Refs.~\cite{Darkfluids1} and \cite{Darkfluids2}. Here, a dark fluid is present which has an inhomogeneous EoS that may depend on the proper evolution of the Universe. This kind of EoS can be derived from the dynamics of an scalar field with some characteristic potential and a variable kinetic term (see \cite{Darkfluids2}), or either it may be seen as the effective EoS corresponding to the addition of various components that fill up our Universe. An EoS of this kind can be written as
\be
p_{ph}=w\rho_{ph} + g(H, \dot{H}, \ddot{H}...;t)\ ,
\label{2.11}
\ee
where $w$ is a constant and $g$ an arbitrary function. The interesting point is that it is possible to specify a function $g$ so that a complete solution of the Friedmann equations is obtained. In this case, our aim is to study a fluid that at present (or in the near future) can behave as phantom; to that purpose  we choose $w=-1$, and the role of the $g$ function will be to determine when exactly the phantom barrier is crossed. Thus, our model will be described by the phantom fluid in (\ref{2.11}) and the function $F(R)$ of (\ref{2.1}), while the matter component can be neglected. As an example, the EoS for the dark fluid is given by
\be
p_{ph}=-\rho_{ph} - \left( \frac{4}{\kappa^2}h'(t)+p_{F(R)}+\rho_{F(R)}\right)\ ,
\label{2.12}
\ee
where the prime denotes  derivative with respect to time, and $p_{F(R)}$ and $\rho_{F(R)}$ are the energy densities defined in (\ref{1.4}). As it is shown, through  the EoS defined in (\ref{2.12}), the dark fluid will contribute to the acceleration of the Universe as a dark energy, and subsequently as a phantom fluid when it crosses the  barrier $w_{ph}<-1$. We thus see that in this model accelerated expansion comes from two contributions, the $F(R)$ term (\ref{2.1}) and the dark fluid one (\ref{2.12}), so that the accelerated expansion stage could cross the phantom barrier if the dark fluid dominates and contributes as a phantom one. The difference with respect to the former model is that in the present case the dark fluid changes its behavior in the course of the expansion history (this will be seen again in an example below). As the EoS given in (\ref{2.12}) can be rewritten in terms of the Ricci scalar $R$, it can be seen as additional terms to our F(R) function, in order to get the transition to the phantom epoch in the context of F(R) gravity. With the EoS (\ref{2.12}), the Friedmann equations can be solved, the following solution being found
\be
H(t)=h(t)\ .
\label{2.13}
\ee
Different solutions can be constructed by specifying the function $h(t)$. We are here interested in those solutions that give rise to a phantom epoch; for those cases the following splitting of the Hubble parameter is relevant
\be
H(t)=\frac{H_0}{t}+\frac{H_1}{t_s-t} \ ,
\label{2.14}
 \ee

where $H_0$, $H_1$ and $t_s$ (Rip time) are positive constants. This function describes a Universe that starts in a singularity, at $t=0$---which may be identified with the Big Bang one---then evolves to a matter dominated epoch, after which an accelerated epoch starts which is dominated by an effective cosmological constant and, finally, the Universe enters into a phantom epoch that will end in a Big Rip singularity. As in the cases above, the condition (\ref{2.2}) remains, that is $F'(R_0)=0$, where $R_0$ is the present value of the Ricci scalar. Fig.~4 is an illustration of how this model works: we see there that the Universe goes through a decelerated epoch until it enters a region where the evolution of its expansion has constant Hubble parameter and the $F(R)$ term behaves as an effective cosmological constant. Finally, at $t_ {ph}$ the Universe enters into a super-accelerated phase which is dominated by the dark fluid of Eq.~(\ref{2.12}), until it eventually reaches the Big Rip singularity at $t=t_s$. The aim of this model is that the crossing phantom barrier takes place very  softly, as it is seen in Fig. 4, due to the two contributions to the acceleration of the Universe expansion.
\\
\\
\\

\begin{figure}[h]
 \centering
 \includegraphics[width=3in,height=3in]{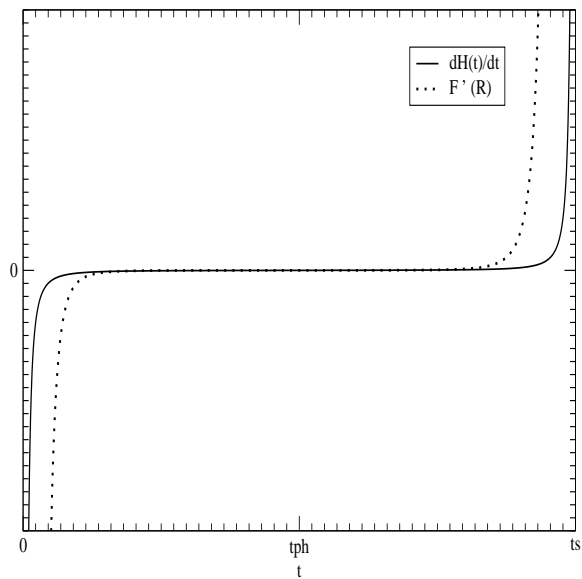}
 \caption{The first derivative of the Hubble parameter is shown here to illustrate the different epochs of the Universe evolution (full curve). Also, the first derivative of the $F(R)$ term with respect to $R$ is represented as a function of time (dotted curve); the constant Hubble parameter region is given by a constant $F(R)$, which plays the role of an effective cosmological constant. Starting at $t=t_{ph}$, the Universe enters into a phantom region dominated by the dark fluid, which ends at $t=t_s$ when the Big Rip singularity takes over.}
 \label{fig:4}
\end{figure}
Alternatively, the evolution of the EoS parameter for the dark fluid (\ref{2.12}) can be written as
\be
w_{ph}=-1-\frac{\frac{4}{\kappa^2}h'(t)+p_{F(R)}+ \rho_{F(R)}}{\frac{6}{\kappa^2}h^2(t)-\rho_{F(R)}}\ ,
\label{2.15}
\ee
and we may look at its asymptotic behavior,
\bea
\text{For} \quad 0<t<<t_{0}\longrightarrow w_{ph}\sim-1+\frac{2(H_0+4)}{3(3-2H^{2}_{0})}\ . \nn
\text{For} \quad t\sim t_0 \longrightarrow w_{ph}\sim-1-\frac{4}{\kappa^2}\dot{H}\ . \nn
\text{For} t_0,t_{ph}<<t<t_s \longrightarrow w_{ph}\sim-1-\frac{2}{9\left(\frac{37H^{2}_1-4H_1+18}{H_1} \right) }\ .
\label{2.16}
\eea

We thus see that, at the starting stages of the Universe, the dark fluid contributes to the deceleration of its expansion.  For $t$ close to the present time, $t_0$, it works as a contribution to an effective cosmological constant  and, after $t=t_{ph}$, it gives rise to the transition to a phantom era of the cosmos, which could actually be taking place nowadays at some regions of it. Our hope is that it could even be observable with specific measurements. Finally, for $t$ close to the Rip time, the Universe  becomes completely dominated by the dark fluid, whose EoS is phantom at that time. This model, which is  able to accurately reproduce the dark energy period, may still be modified in such a way that the epoch which is dominated by the effective cosmological constant, produced by the $F(R)$ term and by the dark fluid contribution, becomes significantly shorter. This is supposed to happen when a matter term is included.

To finish this section, the inclusion of a dark fluid that behaves as a phantom one gives rise to a super-accelerated phase, as compared with the case where just the viable $F(R)$ term contributes. In the two examples here studied, we have proven that, while the $F(R)$ term contributes as an effective cosmological constant, the dark fluid contribution produces the crossing of the phantom barrier, and it continues to dominate until the end of the Universe in a Big Rip singularity. In other words, both the contribution of $F(R)$ and of the phantom fluid are needed. This is very
clearly seen before, and the nice thing is that, because
of this interplay, we have shown the appearance of nice properties of
our model that no purely phantom model could have. Specifically,
for our models above, $F(R)$ gravity together with phantom matter,
the effective $w$ value becomes in fact bigger than -1, so that we
are able to show that $F(R)$ gravity can solve the phantom
problem simply by making the phantom field to appear as a normal one.

\section{Scalar-tensor theories and $F(R)$ gravity with a fluid}

We now turn to the study of the solutions given above in the alternative, and more commonly used, scalar-tensor picture. In Refs.~\cite{F(R)vsScalar2}, \cite{F(R)vsScalar1AndEoS}, \cite{F(R)vsScalar3}, \cite{F(R)vsScalarOriginal}-\cite{F(R)vsScalar5}, it was pointed out that  $F(R)$ gravity can be written in terms of a scalar field---quintessence or phantom like---by redefining the function $F(R)$ with the use of a convenient scalar field and then performing a conformal transformation.  The scalar-tensor  theory thus obtained provides a solution which is characterized by this conformal transformation, whose expression depends on the precise form of the $F(R)$ function. It has been shown that, in general, for any given $F(R)$, the corresponding scalar-tensor theory can in principle be obtained, although the solution is going to be very different from case to case. Also attention has been paid to the reconstruction of $F(R)$ gravity from a given scalar-tensor theory. It is also known (see \cite{F(R)vsScalar2}) that the phantom case in scalar-tensor theory does not allow, in general, the corresponding picture in $F(R)$ gravity. In fact, the conformal transformation becomes complex when the phantom barrier is crossed, and then the resulting $F(R)$ function becomes complex too. To avoid this hindrance, a dark fluid can be used, as in the models of the preceding section, in order to produce the phantom behavior in such a way that the reconstructed $F(R)$ function continues to be {\it real}. This point is important, and will be clearly shown below. Also to be remarked is the fact that scalar-tensor theories, commonly used in cosmology to reproduce dark energy (see for example \cite{Scalar-tensor}), provide cosmological solutions whose stability should be studied in order to demonstrate the validity of the solution found. This has been investigated, e.g., in Ref. \cite{Phantom2} and requires a very deep and careful analysis. For the purpose of the current paper, we will here concentrate in the proof the existence of the solution in the scalar-tensor counterpart, and the corresponding stability study will be left for future work.

We start with the construction of the scalar-tensor theory from $F(R)$ gravity. The action (\ref{1.1}) can be written as
\be
S=\int d^4x \sqrt{-g}\left[P(\phi)R+Q(\phi) + L_m \right]\ ,
\label{3.1}
\ee
which is known as the Jordan frame. Here $F(R)$ has been written in terms of a scalar field. To recover the action (\ref{1.1}) in terms of $F(R)$,  the scalar field equation resulting from the variation of the action (\ref{3.1}) with respect to $\phi$ is used, which can be expressed as follows
\be
P'(\phi)R+Q'(\phi)=0\ ,
\label{3.2}
\ee
where the primes denote derivatives with respect to $\phi$. Then, by solving equation (\ref{3.2}) we get the relation between the scalar field $\phi$ and the Ricci scalar, $\phi=\phi(R)$. In this way, the original $F(R)$ function and the action (\ref{1.1}) are recovered:
\be
R + F(R)= P(\phi(R))R+Q(\phi(R))\ .
\label{3.3}
\ee
Finally, the scalar-tensor picture is obtained by performing a conformal transformation on the action (\ref{3.1}). The relation between both frames is given by
\be
g_{E\mu\nu}= \Omega^2g_{\mu\nu}\ , \quad \text{where} \quad \Omega^2= P(\phi)\ ,
\label{3.4}
\ee
where the subscript $_E$ stands for Einstein frame. A quintessence-like action results in the Einstein frame
\[
S_E=\int d^4x \sqrt{-g_E} \left[R_E -\frac{1}{2} \omega(\phi) d_{\mu}\phi d^{\mu}\phi -U(\phi) +\alpha(\phi) L_{mE} \right]\ ,
\]
where
\be
\omega(\phi)=\frac{12}{P(\phi)}\left( \frac{d\sqrt{P(\phi)}}{d\phi}\right)^2\ , \quad U(\phi)=\frac{Q(\phi)}{P^2(\phi)}\ \quad \text{and} \quad \alpha(\phi)= P(\phi) \ ,
\label{3.5}
\ee
are the kinetic term, the scalar potential and the coupling function, respectively. Hence, by following the steps enumerated above, we can reconstruct the scalar-tensor theory described by the action (\ref{3.5}) for a given $F(R)$ gravity. By redefining the scalar field $\phi=R$, and after combining Eqs.~(\ref{3.2}) and (\ref{3.3}), the form of the two functions $P(\phi)$ and $Q(\phi)$ are found
\be
P(\phi)=1+F'(\phi)\ , \quad Q(\phi)=F'(\phi)\phi -F(\phi) \ .
\label{3.6}
\ee
Hence, for a given solution in the Jordan frame (\ref{3.1}), the solution in the corresponding quintessence/phantom scalar field scenario---i.e., in the Einstein frame (\ref{3.5})---is obtained by the conformal transformation (\ref{3.4}), and it is given by
\be
a_E(t_E)= \left[ 1+F'(\phi(t))\right] ^{1/2}a(t)\ \quad \text{where} \quad dt_E= \left[ 1+F'(\phi(t))\right] ^{1/2}dt\ .
\label{3.7}
\ee
We will be here interested in the phantom case. With that purpose, we analyze the model described in the above section by the $F(R)$ function (\ref{2.1}) and the dark fluid with EoS (\ref{2.12}), and whose solution is (\ref{2.14}). For simplicity, we restrict the reconstruction to the phantom epoch, when the solution can be written as $H(t)\sim\frac{H_1}{t_s-t}$, and $F(R)\sim\frac{\alpha}{\gamma}R^2$.  Using (\ref{3.6}), the function $P(\phi(t))$ takes the form  $P(t)\sim2\frac{\alpha}{\gamma}R^2 =\frac{2\alpha H_1(H_1+1)}{\gamma}\frac{1}{(t_s-t)}$ as a function of time $t$ in the Jordan frame. Then, using (\ref{3.7}), the solution in the Einstein frame is found
\be
t_E=-\sqrt{\frac{2\alpha H_1(H_1+1)}{\gamma}}\ln(t_s-t) \longrightarrow a_E(t_E)= \sqrt{\frac{2\alpha H_1(H_1+1)}{\gamma}} \exp\left[2 \sqrt{\frac{\gamma}{2\alpha H_1(H_1+1)}}t_E\right]\ .
\label{3.8}
\ee
Through the relation between the time coordinates in both frames, we see that while for the Jordan frame there is a Big Rip singularity, at $t=t_s$, this corresponds in the Einstein frame to $t\rightarrow\infty$, so that the singularity is avoided, and there is no phantom epoch there. By analyzing the scale parameter in the Einstein frame, we realize that it describes a de Sitter Universe, while in the Jordan frame the Universe was described by a phantom expansion. As a consequence, we have shown that a phantom Universe in $F(R)$ gravity may be thoroughly reconstructed as a quintessence-like model, where the phantom behavior is lost completely. This has been achieved in a fairly simple example and constitutes an interesting result.

Let us now explore the opposite way. In this case, a phantom scalar-tensor theory is given and it is $F(R)$ gravity which is reconstructed. As was pointed out in Ref.~\cite{F(R)vsScalar2}, when a phantom scalar field is introduced, then the corresponding $F(R)$ function---which is reconstructed, close to the Big Rip singularity, by means of a conformal transformation that deletes the kinetic term for the scalar field---is in general complex. As a consequence, there is no correspondence in modified gravity when a phantom scalar produces a Big Rip singularity. However, in our case we will analyze a scalar-tensor theory which includes a phantom fluid that is responsible for the phantom epoch and for the Big Rip singularity. In this situation a real $F(R)$ gravity will be generically reconstructed, as we are going to see.

The action that describes the scalar-tensor theory is
\be
S_E=\int d^4x \sqrt{-g_E} \left[R_E -\frac{1}{2} \omega(\phi) d_{\mu}\phi d^{\mu}\phi -U(\phi) +\alpha(\phi) L_{phE} \right]\ ,
\label{3.9}
\ee
where $\alpha(\phi)$ is a coupling function and $L_{phE}$ the Lagrangian for the phantom fluid in the Einstein frame. In our case, we consider a phantom fluid with constant EoS, $p_{phE}=w_{ph}\rho_{phE}$, with $w_{ph}<-1$. The Friedmann equations in this frame are written as
\bea
H_E^2=\frac{\kappa^2}{3}\left(\frac{1}{2}\omega(\phi)\dot{\phi}^2+V(\phi)+\alpha(\phi)\rho_{phE} \right)\ , \nn
\dot{H}_E= -\frac{\kappa^2}{2}\left(\omega(\phi)\dot{\phi}^2+ \alpha(\phi)\rho_{phE}(1+w_{ph})\right)\ .
\label{3.10}
\eea
To solve the above equations, it turns out to be very useful to redefine the scalar field as $\phi=t_E$. Then, for a given solution $H_E(t_E)$, the kinetic term for the scalar field can be written as follows
\be
\omega(\phi)=-\frac{\frac{4}{\kappa^2}(\dot{H}_E+3(1+w_{ph})H^2_E)-
(1+w_{ph})V(\phi)}{1-w_{ph}}\ .
\label{3.11}
\ee
To reconstruct $F(R)$ gravity, we perform a conformal transformation that deletes the kinetic term, namely
\be
g_{\mu\nu E}=\Omega^2g_{\mu\nu}\ , \quad \text{where} \quad \Omega^2=\exp\left[\pm \sqrt{\frac{2}{3}}\kappa\int d\phi\sqrt{\omega(\phi)}\right]\ .
\label{3.12}
\ee
Note that for a phantom scalar field, that is defined by a negative kinetic term, the above conformal transformation would be complex, as remarked in Ref.~\cite{F(R)vsScalar2} and as will be shown below. Thus, the reconstructed action would be complex, and no $F(R)$ gravity could be recovered. By means of the above conformal transformation, action (\ref{3.9}) is given by
\be
S=\int d^4x \sqrt{-g} \left[ \frac{\e^{\left[\pm \sqrt{\frac{2}{3}}\kappa\int d\phi\sqrt{\omega(\phi)}\right]}}{2\kappa^2}R-\e^{\left[\pm 2\sqrt{\frac{2}{3}}\kappa\int d\phi\sqrt{\omega(\phi)}\right]} V(\phi) + L_{ph}\right] \ ,
\label{3.13}
\ee
where $L_{ph}$ is the lagrangian for the phantom fluid in the Jordan frame, whose energy-momentum tensor is related with the one in the Einstein frame by $T^{ph}_{\mu\nu}=\Omega^2T^{phE}_{\mu\nu}$,  and where we have chosen a coupling function $\alpha(\phi)=\Omega^{-4}$, for simplicity. By varying now the action (\ref{3.13}) with respect to $\phi$, the scalar field equation is obtained
\be
R=\e^{\left[\pm \sqrt{\frac{2}{3}}\kappa\int d\phi\sqrt{\omega(\phi)}\right]}\left( 4\kappa^2 V(\phi) \mp \sqrt{\frac{6}{\omega(\phi)}}V'(\phi) \right)\ ,
\label{3.13a}
\ee
which can be solved as $\phi=\phi(R)$, so that by rewriting action (\ref{3.13}), the $F(R)$ gravity picture result:
\be
F(R)=\frac{\e^{\left[\pm \sqrt{\frac{2}{3}}\kappa\int d\phi\sqrt{\omega(\phi)}\right]}}{2\kappa^2}R-\e^{\left[\pm 2\sqrt{\frac{2}{3}}\kappa\int d\phi\sqrt{\omega(\phi)}\right]} V(\phi)\ .
\label{3.14}
\ee
Hence, for some coupling quintessence theory described by action (\ref{3.9}), it is indeed possible to obtain a {\it real} $F(R)$ theory, by studying the system in the Jordan frame through the conformal transformation (\ref{3.12}).

To demonstrate this reconstruction explicitly, an example will now be given. As we are interested in the case of a phantom epoch close to the Big Rip singularity, we will start from a solution in the Einstein frame $H_E\sim\frac{1}{t_s-t_E}$, and a scalar potential given by $V(\phi)\sim(t_s-\phi)^n$, with $n>0$. Then, the kinetic term (\ref{3.11}) is written as
\be
\omega(\phi)\sim\frac{-\frac{2}{\kappa^2}(2+3(1+w_{ph}))}{1-
w_{ph}}\frac{1}{(t_s-\phi)^2}\ .
\label{3.15}
\ee
The solution in the Jordan frame is calculated by performing the conformal transformation (\ref{3.12})
\be
(t_s-t_E)=\left[ \left( 1\mp\frac{k}{2}\right)t \right]^{1/(1\mp k/2)}\ \longrightarrow \quad a(t)\sim \left[ \left( 1\mp\frac{k}{2}\right)t \right]^{-\frac{1\pm k}{1\mp k}}\ ,
\label{3.16}
\ee
where $k=\sqrt{-\frac{8(1+3(1+w_ph))}{3(1-w_{ph})}}$ and $w_{ph}<-1$. Note that, in this case, we can construct two different solutions, and correspondingly two different $F(R)$ models, depending on the sign selected in Eq.~(\ref{3.12}). It is easy to see that the Big Rip singularity is thereby transformed, depending on the case, into an initial singularity (+), or into an infinity singularity(-). By using (\ref{3.13a}) and (\ref{3.14}), the following function $F(R)$ is recovered
\be
F(R)\sim \frac{(\frac{R}{4\kappa^2})^{\frac{\pm k}{n\pm k}}}{2\kappa^2}R-\left( \frac{R}{4\kappa^2}\right)^{\frac{n\pm 2k}{n\pm k}}\ .
\label{3.17}
\ee
To summarize, we have here shown, in an explicit manner, that an $F(R)$ theory can be actually constructed from a phantom model in a scalar-tensor theory  where the scalar field does not behave as a phantom one (in that case the action for $F(R)$  would become complex). Moreover, very interesting $F(R)$ models crossing the phantom divide can also be constructed explicitly \cite{F(R)vsScalar3}. The above reconstruction procedure, where we have taken the $F(R)$ function of (\ref{2.1}), can be generalized to other types of modified gravity models (for recent reviews see \cite{f(R)review1} and \cite{f(R)review2}).

\section{Discussion}

We have seen in this paper that the $F(R)$ model given by (\ref{2.1}), and where  the $F(R)$ fluid behaves as an effective cosmological constant, is able to reproduce the same behavior, at present time, as the $\Lambda$CDM model. On the other hand, this model gives rise to the accelerated expansion of the inflation epoch too, so that the natural next step to undertake with those models should be to study the complete cosmological history, in particular (what is very important) the explicit details for a grateful exit from inflation, what would demonstrate their actual potential. What is more, $F(R)$ functions of this kind might even lead to a solution of the cosmological constant problem, by involving a relaxation mechanism of the cosmological constant, as was indicated in \cite{CosmoProblem} (see also the very recent paper \cite{bss1}). The effective cosmological constant thus obtained could eventually adjust to its precise observable value. This issue is central and deserves further investigation.

We have studied the behavior of the cosmological evolution when a phantom fluid is introduced that contributes to the accelerated expansion of the universe. The $F(R)$ function we have considered is the one given in (\ref{1.9}), which contributes as an effective cosmological constant at present time. The motivation for this case study comes, as it will be further explored, from the difficulty one encounters to construct {\it viable} $F(R)$ functions which produce phantom epochs, and this has a representation in the scalar-tensor picture in the Einstein frame. As recent observational data suggest (see \cite{obsData1} and \cite{obsData2}), the effective EoS parameter for dark energy is around $-1$, so that the phantom case is allowed---and actually favored by recent, extensive observations.

We thus have seen that, at the early stages of the universe history, the dark fluid contributes to the deceleration of its expansion.  For $t$ close to present time, $t_0$, it works as a contribution to an effective cosmological constant and, after $t=t_{ph}$, it gives rise to the transition to a phantom era of the universe, which could actually be taking place right now in some regions of it. Our hope is that it could be actually observable. Finally, for $t$ close to the Rip time, the Universe  becomes completely dominated by the dark fluid, whose EoS is phantom-like at that time. This model, which is able to reproduce the dark energy period quite precisely, may still be modified in such a way that the epoch dominated by the effective cosmological constant, produced by the $F(R)$ term and by the dark fluid contribution, becomes significantly shorter. This is the case when a matter term is included.

The inclusion of a dark fluid with phantom behavior gives rise to a super-accelerated phase, as compared with the case where just the viable $F(R)$ term contributes. In the two examples investigated in the paper we have proven that, while the $F(R)$ term contributes as an effective cosmological constant, the dark fluid term produces the crossing of the phantom barrier, and it continues to dominate until the end of the universe in a Big Rip singularity. It is for this reason that the contribution of
$F(R)$ and of the phantom fluid are both {\it fundamental}. This has been
clearly explained in the paper, and the nice thing is that, because
of this interplay, we have shown the appearance of very nice properties of
our model that no purely phantom model could have. This eliminates
in fact some of the problems traditionally associated with phantom
models and makes this study specially interesting. In particular,
for some of our models of $F(R)$ gravity together with phantom matter,
the effective $w$ value becomes in fact bigger than -1, so that we
were able to show that $F(R)$ gravity can solve the phantom
problem simply by making the phantom field to appear as a normal field.

To summarize, we have here shown, in an explicit manner, that an $F(R)$ theory can indeed be constructed from a phantom model in a scalar-tensor theory in which the scalar field does not behave as a phantom one (in the latter case the action for $F(R)$  would be complex). Moreover, very promising $F(R)$ models which cross the phantom divide can be constructed explicitly (see also \cite{F(R)vsScalar3}). The above reconstruction procedure, where we have taken (\ref{2.1}) for the $F(R)$ function, can be generalized to other classes of modified gravity models. \bigskip

\noindent {\bf Acknowledgments.} We thank Sergei Odintsov for useful discussion and would like to congratulate him on the occasion of his 50th Birthday. We also thank the referee of a previous version of this paper for comments and criticisms that led to its improvement. This investigation has been supported in part by MICINN (Spain), project FIS2006-02842, and by AGAUR (Generalitat de  Ca\-ta\-lu\-nya), contract 2005SGR-00790 and grant DGR2008BE1-00180. DSG acknowledges a grant from MICINN. Part of EE's research was performed while on leave at Department of Physics and Astronomy, Dartmouth College, 6127 Wilder Laboratory, Hanover, NH 03755, USA.

\end{document}